\documentclass[floats,preprint,aps,pra]{revtex4}
\usepackage{graphicx,amsmath,amsthm,amssymb}
\usepackage{amsfonts}
\usepackage{amsmath}
\usepackage{amssymb}
\usepackage{graphicx}

\input{tcilatex}

\begin{document}

\title{Causal vs. Noncausal Description of Nonlinear Wave Mixing; Resolving
the Damping-Sign Controversy}
\author{Shaul Mukamel}
\affiliation{Department of Chemistry, University of California, Irvine, CA 92697}
\date{\today}

\begin{abstract}
Frequency-domain nonlinear wave mixing processes may be described either
using response functions whereby the signal is generated after all
interactions with the incoming fields, or in terms of scattering amplitudes
where all fields are treated symetrically with no specific time ordering. \
Closed Green's function expressions derived for the two types of signals
have different analytical properties. \ The recent controversy regarding the
sign of radiative damping in the linear (Kramers Heisenberg) formula is put
in a broader context.

Submitted to Phys. Rev. A (Rapid communication)

PACS Indices

42.65.An, 32.80.-t, 42.50.ct, 32.70.Jz
\end{abstract}

\maketitle


\section{Introduction}

A lively debate is currently going on with regard to the correct sign that
should be introduced in the Kramers-Heisenberg expression for the (linear)
optical polarizability. \ Both "opposite sign" and a "constant sign"
formulas have been derived by various authors. \ Perturbative QED
calculations have been carried out to include radiative damping [1-8]. \ In
an insightful recent article Bialyncki-Birula and Sowinski\ [9] had pointed
out that this issue is fundamentally connected with the linear response vs.
the scattering points of view for Rayleigh scattering. \ In this letter we
extend this argument to nonlinear wave mixing of arbitrary order. \ Closed
formal expressions are derived that reveal the analytical properties of both
retarded (response) and non retarded (scattering) signals. \ The linear
response results are recovered to lowest order. \ Our microscopic deviation
could serve as the starting point for a full QED calculation. \ However,
this is not required in order to pinpoint the analytical properties which
should hold for other types of nonradiative damping as well. \ 

The following derivation applies for processes of arbitrary order. \
However, for clarity we focus on four wave mixing. \ Consider a system
(atom, molecule) interacting with four modes of the radiation field. \ The
states of the system will be denoted a,b,c.... \ . \ The j'th mode of the
radiation field has a frequency $\omega _{j}$ and initial occupation number $%
n_{j}$. \ To simplify the notation we hereafter consider sum frequency
generation whereby $\omega _{4}=\omega _{1}+\omega _{2}+\omega _{3}$,
however the results can be easily extended to any combination of frequencies 
$\omega _{4}=\pm \omega _{1}\pm \omega _{2}+\pm \omega _{3}$. \ For the
process of interest the initial state of the system + field is $\mid i>=\mid
a,$ $n_{1}n_{2}n_{3}n_{4}>,$ and the final state is $\mid f>=\mid a,$ $%
n_{1}-1,n_{2}-1,n_{3}-1,n_{4}+1>$, with energies $E_{i}$ and $E_{f}$
respectively.

The coupling of an atom located at point $\mathbf{r}$ with the radiation
field is 
\begin{equation}
H_{int}=-V\text{\textbf{.}}E(\mathbf{r},t).  \label{aa}
\end{equation}

We shall divide the electric field into its positive and negative frequency
components

\begin{equation}
E(\mathbf{r},t)=\varepsilon (\mathbf{r},t)+\varepsilon ^{\dagger }(\mathbf{r}%
,t)
\end{equation}

\begin{equation}
\varepsilon (\mathbf{r},t)=\sum_{j}(\frac{2\pi \hbar \omega _{j}}{\Omega }%
)^{1/2}\exp (ik_{j}\mathbf{r}-i\omega _{j}t)a_{j}
\end{equation}%
$a_{j}$ is the photon annihilation operator of mode $j$,\ V is the dipole
operator and $\Omega $ is the quantization volume.

We first assume that the system is prepared in a nonequilibrium steady state
with the four fields and consider the entire process as a single four-photon
scattering event. \ This is a \textit{non causal} process whereby all four
modes are treated along the same footing. \ This implies that the
interaction with model 4 need not be the last, and all time orderings are
allowed and should be summed over. \ The process is described by the S
matrix element%
\begin{equation}
S_{fi}=A_{fi}T_{fi}(E_{i})\delta (E_{i}-E_{f}),
\end{equation}

where $T_{fi}$ is the matrix element of the T matrix T=V+V G(E) V, and%
\begin{equation}
G(E)=\frac{1}{E-H+i\varepsilon }
\end{equation}%
is the retarded Green's function.\ \ The scattering amplitude for this
process is given by 
\begin{equation}
S_{fi}^{(4)}=A_{fi}\sum_{p_{4}}  \label{ab}
\end{equation}

\begin{equation*}
<a\mid VG(E_{a}+\omega _{1}+\omega _{2}+\omega _{3})VG(E_{a}+\omega
_{1}+\omega _{2})VG(E_{a}+\omega _{1})V\mid a>\delta (\omega _{1}+\omega
_{2}+\omega _{3}-\omega _{4})
\end{equation*}

where

\begin{equation}
A_{fi}=(\frac{2\pi \hbar }{\Omega })^{2}\sqrt{n_{1}n_{2}n_{3}(n_{4}+1)}\sqrt{%
\omega _{1}\omega _{2}\omega _{3}\omega _{4}}.
\end{equation}

$\sum_{p_{4}}$denotes the sum over all 4! permutations of $\omega
_{1},\omega _{2},\omega _{3},$ and -$\omega _{4}$. The sign convention for $%
\omega _{j}$ in eq.6 is as follows: an absorbed photon gives $+\omega _{j}$
whereas an emitted photon gives $-\omega _{j}$. \ In the process considered
here, $\omega _{1},\omega _{2}$ and $\omega _{3}$ are absorbed and $\omega
_{4}$ is emitted. \ Other processes can be calculated by simply changing the
signs of $\omega _{j}$,as warranted in each case. \ H is the Hamiltonian for
the atom + all modes of the radiation field excluding the 4 modes of
interest, since those were treated perturbatively.

We next turn to the standard semiclassical description of four wave mixing
whereby the system first interacts with modes 1, 2 and 3 to generate the
third-order polarization which then serves as a source for mode 4 [10]. \
This is a \textit{causal} response where mode 4 is special since its
interaction with the system must be the very last. \ We shall calculate the
n'th order polarization as the expectation value of V [11]%
\begin{equation}
P^{(n)}=\sum_{m=0}^{n}<\Psi ^{(m)}\mid V\mid \Psi ^{(n-m)}>,
\end{equation}%
where $\Psi ^{(m)}$ is the perturbed wavefunction to $m^{\prime }th$ order
in $V$. \ The $m^{\prime }th$ term represents ($n-m)^{\prime }th$ order for
the ket and $m^{\prime }th$ order for the bra. \ Overall there are $(n+1)$
terms. \ For $n=3$ we get four terms corresponding to $m=0,1,2,3$
respectively in eq.(8). \ We then get $P^{(3)}=A_{fi}$ $\chi _{fi}^{(3)}$
where

\begin{eqnarray}
\chi _{fi}^{(3)} &=&\sum_{p_{3}} \\
&<&a\mid VG(E_{a}+\omega _{1}+\omega _{2}+\omega _{3})VG(E_{a}+\omega
_{1}+\omega _{2})VG(E_{a}+\omega _{1})V\mid a>  \notag \\
&<&a\mid VG^{\dagger }(E_{a}+\omega _{1}+\omega _{2}-\omega
_{4})VG(E_{a}+\omega _{1}+\omega _{2})VG(E_{a}+\omega _{1})V\mid a>  \notag
\\
+ &<&a\mid VG^{\dagger }(\omega _{1}-\omega _{4}+\omega _{2})VG^{^{\dagger
}}(E_{a}+\omega _{1}-\omega _{4})VG(E_{a}+\omega _{1})V\mid a>  \notag \\
+ &<&a\mid VG^{\dagger }(E_{a}-\omega _{4}+\omega _{1}+\omega
_{2})VG^{^{\dagger }}(E_{a}-\omega _{4}+\omega _{1})VG^{^{\dagger
}}(E_{a}-\omega _{4})V\mid a>  \notag \\
&&\delta (\omega _{1}+\omega _{2}+\omega _{3}-\omega _{4}),  \notag
\end{eqnarray}

is the susceptibility and 
\begin{equation}
G^{^{\dagger }}(E)=\frac{1}{E-H-i\varepsilon },
\end{equation}%
is the advanced Green's function. \ $\sum_{p_{3}}$ denotes the sum over all
3! permutations of the \textit{incoming} modes $\omega _{1},\omega _{2}$ and 
$\omega _{3}.$ \ In this expression the sign conversion of $\omega _{j}$ is
the same as in eq. (6) (+sign for absorbed photons, - for emitted). \ Each
coupling with the ket is accompanied by a retarded Green's function $G$,
whereas an advanced Green's function $G^{\dagger }$ is accompanied by bra
interactions. Eq. (8) will thus yield $(n-m)$ $G$ and $m$ $G^{\dagger }$
factors. \ An analogous expression was obtained recently using
superoperators in Liouville Space. [12] \ The present Hilbert Space form is
more suitable for comparison with the scattering amplitudes.

The extension of Eqs.(6) and (9) to arbitrary order is straightforward. \ $%
S^{(n+1)}$ will have a single basic term with $(n+1)$ V factors, and $n$
retarded Green's functions $G$ with arguments $E_{a}+\omega _{1}$, ...$%
E_{a}+\omega _{1}+\omega _{2}+...+\omega _{n.}.$ In addition, it contains a
sum over the $(n+1)!$ permutations of $\omega _{1},$..., $\omega _{n+1}.$
Each $\omega _{j}$ may be changed to $-\omega _{j}$ to describe different
processes. \ $\chi ^{(n)}$ has $(n+1)$ basic terms each containing $(n+1)$ V
factors, and a $(G^{+})^{m}(G)^{n-m}$ factor with $m=0,...n$ (See Eq.8). The
retarded Green's functions will only depend on the incoming frequencies $%
\omega _{1,}$...,$\omega _{n}.$ \ All advanced Green's functions also depend
on the signal frequency $-\omega _{n+1}$. \ Each basic term yields $n!$
terms upon the permutation over $\omega _{1}$.. $\omega _{n}$. Altogether
both $S^{(n+1)}$ and $\chi ^{(n)}$ contain $(n+1)$! terms. \ For $S^{(n+1)}$
these come from the $(n+1)$! permutations of the $(n+1)$ frequencies over a
single term. \ $\chi ^{(n)}$ has $(n+1)$ basic terms, each containing $n$! \
permutations of the $n$ "incoming" frequencies. \ Unlike $\chi ^{(n)}$, $%
S^{(n+1)}$ is symmetric with respect to all $(n+1)$ modes; the interaction
with the signal field $\ \omega _{n+1\text{ }}$need not be chronologically
the last. $\ S^{(n+1)}$ only contains retarded Green's functions and all
propagations are forward in time. \ The $(n+1)$! permutations take care of
the possible time orderings of interactions with the various modes. \ $\chi
^{(n)},$ in contrast, depends on both retarded and advanced Green's
functions which correspond to forward and backward propagations respectively
along the Keldysh Schwinger loop [13-15].

Both expressions (6) or (9), can serve as a starting point for a full QED
perturbative calculation, where the coupling with all modes of the
electromagnetic field is included in the Hamiltonian H. This will result in
damping terms. \ However, the analytical properties of the two signals are
completely determined by eqs (6) or (9), and will be invariant to the level
of approximation used for the radiative damping. \ When all frequencies are
tuned off resonance, we can neglect the $\pm i\varepsilon $ terms in the
Green's functions, setting $G=G^{^{\dagger }}$. \ The causal and non causal
expressions are then identical.

The ongoing damping sign controversy was restricted to the linear response
of a two level system with ground state a and an excited state b [1-9]. \
This may be immediately resolved by our general formulation. \ For Rayleigh
(elastic) scattering we have one absorbed and one emitted photon with the
same frequency. \ We thus set $\omega _{1}=\omega $ and $\omega _{2}=-\omega
.$ The second order analogue of eq.(6) reads%
\begin{equation}
S_{fi}^{(2)}\sim <a\mid VG(E_{a}+\omega )V\mid a>+<a\mid VG(E_{a}-\omega
)V\mid a>
\end{equation}

This is obtained from a single term + 2 permutations of $\omega $ and $%
-\omega $. \ Writing the matrix elements explicitly for the system
Hamiltonian (neglecting coupling with other radiation modes), this gives

\begin{equation}
S_{fi}^{(2)}\sim \mid V_{ab}\mid ^{2}\left[ \frac{1}{E_{a}+\omega
-E_{b}+i\varepsilon }+\frac{1}{E_{a}-\omega -E_{b}+i\varepsilon }\right] .
\end{equation}

This is known as the "constant sign" prescription (both terms have a $%
+i\varepsilon $ factor) [1-9]. The linear response expression similar to
eq.(9) gives on the other hand%
\begin{equation}
\chi _{fi}^{(1)}\sim <a\mid VG(E_{a}+\omega )V\mid a>+<a\mid VG^{^{\dagger
}}(E_{a}-\omega )V\mid a>
\end{equation}

This comes from two basic terms with \textit{no} permutation. \ (there is
only one incoming field which is absorbed and has frequency $\omega $). \
Taking the matrix elements we recover the "opposite sign" prescription%
\begin{equation}
\chi _{fi}^{(1)}\sim \mid V_{ab}\mid ^{2}\left[ \frac{1}{E_{a}+\omega
-E_{b}+i\varepsilon }+\frac{1}{E_{a}-\omega -E_{b}-i\varepsilon }\right] .
\end{equation}

The origin of the different damping signs of $S^{(2)}$ and $\chi ^{(1)}$ had
been clearly pointed out by Bialynicki-Birulya and Sowinski [9] who had
further carried out a fourth order QED calculation of dampling using both
expressions. \ The present results extend these arguments to nonlinear
processes of arbitrary order. It should be emphasized that the fundamental
principle of causality must always hold and is never in doubt. \ The terms
"causal" and "noncausal" in this letter refer to different observables. \
Traditional semiclassical formulation of nonlinear optics imposes a certain
time ordering by singling out one of the fields. \ This response is causal.
\ The scattering description is noncausal since it allows for arbitrary time
ordering of interactions with the various fields.

The support of the National Science Foundation (Grant No. CHE-0446555) and
NIRT (Grant No. EEC0303389) is gratefully acknowledged. \ I wish to thank
Professor Bialynicki Birula for useful discussions.

\bigskip

\textbf{References}

1.\qquad D.L. Andrews, S. Naguleswaran, and G.E. Stedman, Phys. Rev. A 57,
4925 (1998).

2.\qquad A.D. Buckingham and P. Fischer, Phys.Rev. A 61, 35801 (2000).

3.\qquad A.D. Buckingham and P. Fischer, Phys.Rev. A 63, 47802 (2001).

4.\qquad G.E. Stedman, S. Naguleswaran, D.L. Andrews, and L.C. Davila
Romero, Phys. Rev. A 63, 47801 (2003).

5.\qquad D.L. Andrews, L.C. Davila Romero and G.E. Stedman, Phys. Rev. A 67,
55801 (2003).

6.\qquad P.W. Milonni and R.W. Boyd, Phys. Rev. A 69, 23814 (2004).

7.\qquad P.R. Berman, R.W. Boyd and P.W. Milonni, Phys. Rev. A 74, 53816
(2006).

8.\qquad R. Loudon and S.M. Barnett. J. Phys. B: At. Mol. Opt. Phys. 39,
S555 (2006).

9.\qquad I. Bialynicki-Birula and T. Sowinski, Quantum electrodynamics of
qubits, ArXiv:0705.2121v1.

10.\qquad N. Bloembergen, \textit{Nonlinear optics} (Benjamin, New York,
1965).

11.\qquad S. Mukamel, \textit{Principles of Nonlinear Spectroscopy} (Oxford
University Press, New York, 1995).

12.\qquad S. Mukamel, \textit{Partially-Time-Ordered Keldysh-Loop Expansion
of Coherent Nonlinear Optical Susceptibilities}, Submitted to Phys. Rev. A.\
(2007).

13.\qquad L. V. Keldysh, Sov. Phys. JETP \textbf{20}, 1018 (1965).

14.\qquad R. Mills, \textit{Propagators for many-particle systems; an
elementary treatment} (New York, Gordon and Breach 1969).

15.\qquad H. Haug and A-P. Jauho, \textit{Quantum Kinetics in Transport and
Optics of Semiconductors} (Springer-Verlag, Berlin,\ Heidelberg, 1996).

\end{document}